# Coulomb-Volkov approach of ionization by extreme ultraviolet laser pulses in the subfemtosecond regime

*Duchateau G., Cormier E. and Gayet R..*

CELIA[1], Université Bordeaux 1, 351 Cours de la Libération, 33405 Talence Cedex, France.

01/10/02

[1] Centre Lasers Intenses et Applications (UMR 5107, Unité Mixte de Recherche CNRS-Université Bordeaux 1) ; http://www.celia.u-bordeaux.fr

**Abstract**

In conditions where the interaction betweeen an atom and a short high-frequency extreme ultraviolet laser pulse is a perturbation, we show that a simple theoretical approach, based on Coulomb-Volkov-type states, can make reliable predictions for ionization. To avoid any additional approximation, we consider here a standard case : the ionization of hydrogen atoms initially in their ground state. For any field parameter, we show that the method provides accurate energy spectra of ejected electrons, including many above threshold ionization peaks, as long as the two following conditions are simultaneously fulfilled : (i) the photon energy is greater than or equal to the ionization potential ; (ii) the ionization process is not saturated. Thus, ionization of atoms or molecules by the high-harmonics laser pulses which are generated at present may be addressed through this Coulomb-Volkov treatment.

## I. Introduction

In the last months, low intensity X-UV laser pulses of a few hundreds of attoseconds have been produced by harmonics generation techniques [1]. Such pulses are promising powerful tools to investigate thoroughly the dynamics of various elementary processes (e.g., electronic evolution of microscopic systems). Further, the energy of high-harmonic photons is often high enough to ionize atomic or molecular systems with the absorption of a single photon. It is at variance with infrared laser pulses, where both a simultaneous absorption of many photons and a high laser intensity are required to achieve ionization with a reasonable probability. Furthermore, with very short X-UV laser pulses, the ionization of a microscopic system takes place in a time interval much shorter than any relaxation period of the system. Untill now, most theoretical approaches of ionization were made in a context of both much longer pulse durations and much lower photon energies. Indeed, full numerical treatments of the time dependent Schrödinger equation are powerful ways to investigate the ionization of one or two-electron systems by subfemtosecond X-UV laser pulses. However, running them involves intensive calculations [2]. Therefore, the studies of future experiments call for adapted theoretical approximations that could provide easy reliable predictions of ionization by short X-UV laser pulses. In this new context, we examine an approach that was introduced in the seventies [3]. In order to avoid any further complication, we consider here the classic example of the ionization of a hydrogen atom initially in its ground state. A code implementing a full 3D numerical approach, hereafter referred to as TDSE [2], is used to indicate the domain where this approximation is applicable.

In previous papers [4-6], we introduced a non-perturbative approach of atom ionization by ultra-short electromagnetic pulses. It was based on Coulomb-Volkov (CV) states [3], which are used in the framework of the sudden approximation. Compared to TDSE calculations, this approach, hereafter referred to as CV1, appears to be a very simple but powerful tool to study ionization by an external electric field, no matter what the strength of the perturbation is, when three conditions are simultaneously fulfilled : (i) the pulse duration is smaller than or comparable to the initial orbital period of the electron, (ii) all this time the electric field does not perform more than two oscillations, and (iii) its net integral over this time is not zero (DC component of the electric field that shows here mainly a classical aspect). This regime may be called a *collisional regime* because the electromagnetic pulse presents similarities to the perturbation produced by a fast impinging ion. CV1 predictions are shown to be all the better that the electric field amplitude is high, the initial quantum number is large and the pulse duration is short. Thus, it should be well adapted to study the ionization by impact of swift ions, more especially by fast multicharged ions. Although CV1 cannot be used to make predictions with genuine laser pulses (for which the DC electric field component is zero), the previous studies indicate that, during a short enough time, the exact state of the interacting system is well described by CV wavefunctions.

High-harmonics subfemtosecond pulses fit in with a quite different context. They are very short and their intensity is quite small (see e.g., ref. [1]). Thus, investigations may be restricted to conditions where the laser field is a perturbation. Further, a pulse always contains enough field oscillations to appear as a photon field, whose spectrum may be quite broad, depending upon the pulse length. Now, the field exhibits mainly its quantum aspect. With the absorption by a hydrogen atom of a photon, whose energy is high enough to ionize the atom, the electron is rapidly transferred to the continuum. Thus, a Coulomb-Volkov approach restricted to perturbation conditions may be imagined since a Coulomb-Volkov state is a good description of the interacting system as long as the time of energy transfer is small compared to any characteristic atomic period (e.g., the orbital period) [6].

The interest of this type of approach comes from the fact that the Coulomb-Volkov phase offers all possibilities of simultaneous exchange of many photons. In addition, the Coulomb influence of the nucleus *before* and *after* interaction is kept, thus preserving the asymptotic behaviours. A similar approach has already been introduced a long time ago by Jain and Tzoar [3]. The method was examined in various situations (stationary laser beams, low frequency photons) which differ completely from the physical conditions of ultra-short X-UV laser pulses (see, e.g., [7,8] and references therein). In the present paper, we investigate to which extent a similar perturbation Coulomb-Volkov approach may be employed to predict single ionization of atoms or molecules by high-frequency photons of a short laser pulse. It is worth noting that *our study does not include backscattering*, thus excluding all further processes such as higher harmonics emission or high-energy ATI.

In section II, the perturbative Coulomb-Volkov theory of ionization, hereafter referred to as CV2$^-$, is briefly described in the context of short X-UV laser pulses. In section III, it is applied in conditions that can be achieved actually with present-day high-harmonics radiation. CV2$^-$ energy spectra of electrons ejected from hydrogen atoms initially in their ground state are compared to TDSE predictions. Conclusions are drawn in section IV.

Atomic units are used throughout unless otherwise stated.

## II. Theory

In non-relativistic conditions, the wavefunction $\Psi(\vec{r},t)$ of a monoelectronic system interacting with an external electromagnetic field $\vec{F}(\vec{r},t)$, that is assumed to be almost uniform in a large region around the atom at a given time *t* (dipole approximation), is given by the time-dependent Schrödinger equation :

$$i\frac{\partial\Psi(\vec{r},t)}{\partial t} = \left[ H_a + \vec{r}.\vec{F}(t)\right]\Psi(\vec{r},t)$$
$$H_a = -\frac{\nabla^2}{2} + V_a(\vec{r}) \qquad (1)$$

where $\vec{r}$ gives the position of the electron with respect to the nucleus identified with the centre-of-mass. $\vec{F}(t)$ is the external field at the atom. $V_a(\vec{r})$ represents the interaction between the electron and the rest of the target. With a hydrogen-like target of nuclear charge *Z*, it is simply :

$$V_a(\vec{r}) = -\frac{Z}{r} \qquad (2)$$

In what follows, the study is made for a hydrogen atom initially in the ground state. However, the formalism can be extended to atoms or ions with a single valence electron using a procedure similar to the one of ref. [9] for alkali-metal atoms. Thus, the field-free initial state is :

$$\phi_i(\vec{r},t) = \varphi_i(\vec{r})\ \exp\left(-i\varepsilon_i t\right) \qquad (3)$$

where $\varepsilon_i = -0.5$ is the energy of the ground state $\varphi_i(\vec{r})$ which is :

$$\varphi_i(\vec{r}) = \frac{e^{-r}}{\sqrt{\pi}} \qquad (4)$$

The unperturbed final continuum state $\phi_f^-(\vec{r},t)$ is the ingoing regular Coulomb wavefunction:

$$\phi_f^-(\vec{r},t) = \varphi_f^-(\vec{r})\ \exp\left(-i\varepsilon_f t\right) \qquad (5)$$

where $\varphi_f^-(\vec{r})$ is a continuum state of hydrogen normalized to $\delta\left(\vec{k}-\vec{k}'\right)$; it is explicitly :

$$\varphi_f^-(\vec{r}) = (2\pi)^{-\frac{3}{2}}\exp\left(+\frac{\pi\nu}{2}\right)\Gamma(1+i\nu)\ \exp\left(i\vec{k}.\vec{r}\right)\ {}_1F_1(-i\nu;1;-ikr-i\vec{k}.\vec{r}) \qquad (6)$$

where $\vec{k}$ is the electron momentum ; $\varepsilon_f = k^2/2$ is the eigenenergy of $\varphi_f^-(\vec{r})$ and $\nu = 1/k$. Both $\varphi_i(\vec{r})$ and $\varphi_f^-(\vec{r})$ are eigenstates of the field-free hamiltonian $H_a$.

The finite pulse duration is featured through a sine -square envelope. Thus, in the vicinity of the atom, the external field reads :

$$\begin{cases} \vec{F}(t) = \vec{F}_0 \sin(\omega t + \varphi) \sin^2\left(\dfrac{\pi t}{\tau}\right) & \text{when} \quad t \in [0, \tau] \\ \vec{F}(t) = \vec{0} & \text{elsewhere} \end{cases} \tag{7}$$

where $\tau$ is the total duration of the pulse. In what follows, we choose $\omega = 0.855\, a.u.$ in order to have a photon energy that corresponds to the average high -harmonics energy reported in ref.[1] (15[th] harmonics). Although it is not of great importance when many oscillations are performed within $[0, \tau]$, all calculations are made as in paper [4] with a time -symmetric pulse, which implies $\varphi = \frac{\pi}{2} - \omega \frac{\tau}{2}$. The electric field of the laser is derived from a vector potential $\vec{A}(t)$ that reads :

$$\vec{A}(t) = \vec{A}(t_0) - \int_{t_0}^{t} dt' \vec{F}(t') \tag{8}$$

With the final state $\phi_f^-(\vec{r}, t)$, one builds an ingoing Coulomb-Volkov wave function $\chi_f^-(\vec{r}, t)$. According to [3,4], it is :

$$\begin{cases} \chi_f^-(\vec{r}, t) = \phi_f^-(\vec{r}, t)\; L^-(\vec{r}, t) \\ L^-(\vec{r}, t) = \exp\left\{ i\vec{A}^-(t).\vec{r} \; - \; i\vec{k} \cdot \displaystyle\int_{\tau}^{t} dt' \vec{A}^-(t') \; - \; \frac{i}{2} \int_{\tau}^{t} dt' \vec{A}^{-2}(t') \right\} \end{cases} \tag{9}$$

where $\vec{A}^-(t)$ is the variation of $\vec{A}(t)$ over the time interval $[\tau, t]$, i.e. :

$$\vec{A}^-(t) = \vec{A}(t) - \vec{A}(\tau) = -\int_{\tau}^{t} \vec{F}(t)\, dt \tag{10}$$

In previous papers [4,6], it is shown that, within an interaction time shorter than the initial orbital period, a Coulomb -Volkov wavefunction gives a good representation of the interacting system.

In the Schrödinger picture, the transition amplitude from the state $i$ at $t \to -\infty$ to the final state $f$ at $t \to +\infty$ may be evaluated at any time $t$; it is :

$$T_{fi} = \left\langle \Psi_f^-(t) \,\middle|\, \Psi_i^+(t) \right\rangle \tag{11}$$

where $\Psi_f^-(\vec{r},t)$ and $\Psi_i^+(\vec{r},t)$ are the exact solutions of the equation (1) subject to the asymptotic conditions :

$$\Psi_f^-(\vec{r},t) \underset{t\to+\infty}{\longrightarrow} \phi_f^-(\vec{r},t) \tag{12a}$$

$$\Psi_i^+(\vec{r},t) \underset{t\to-\infty}{\longrightarrow} \phi_i(\vec{r},t) \tag{12b}$$

In order to use the Coulomb -Volkov wave function $\chi_f^-(\vec{r},t)$, calculations are made with the so-called *prior* form of the transition amplitude that is :

$$T_{fi}^- = \lim_{t\to-\infty} \left\langle \Psi_f^-(t) \middle| \Psi_i^+(t) \right\rangle = \lim_{t\to-\infty} \left\langle \Psi_f^-(t) \middle| \phi_i(t) \right\rangle \tag{13}$$

according to (12b) and because $\phi_i(\vec{r},t)$ and $\phi_f^-(\vec{r},t)$ are orthogonal, one may write :

$$T_{fi}^- = \lim_{t\to-\infty} \left\langle \Psi_f^-(t) \middle| \phi_i(t) \right\rangle - \lim_{t\to+\infty} \left\langle \Psi_f^-(t) \middle| \phi_i(t) \right\rangle = \int_{+\infty}^{-\infty} dt \frac{\partial}{\partial t} \left\langle \Psi_f^-(t) \middle| \phi_i(t) \right\rangle \tag{14}$$

After a standard easy algebra, the expression (14) may be transformed into :

$$T_{fi}^- = -i \int_0^\tau dt \left\langle \Psi_f^-(t) \middle| \vec{r} \cdot \vec{F}(t) \middle| \phi_i(t) \right\rangle \tag{15}$$

In perturbative conditions, one may substitute $\chi_f^-(\vec{r},t)$ to $\Psi_f^-(\vec{r},t)$ in (12). Then, according to expressions (3), (5) and (9), $T_{fi}^-$ may be written as :

$$\begin{aligned} T_{fi}^- \cong T_{fi}^{CV2-} = -i \int_0^\tau dt \exp\left\{ i \left( \frac{k^2}{2} - \varepsilon_i \right) t + i\,\vec{k}.\int_\tau^t dt' \vec{A}^-(t') + \frac{i}{2} \int_\tau^t dt' \vec{A}^{-2}(t') \right\} \\ \times \int d\vec{r}\, \varphi_f^{-*}(\vec{r}) \exp\left( -i\vec{A}^-(t).\vec{r} \right) \vec{r}.\vec{F}(t)\, \varphi_i(\vec{r}) \end{aligned} \tag{16}$$

Let us introduce the useful functions :

$$h^-(t) = i\left( \frac{k^2}{2} - \varepsilon_i \right) + i\,\vec{k}.\vec{A}^-(t) + \frac{i}{2} \vec{A}^{-2}(t) \tag{17}$$

$$f^-(t) = \exp\left\{ \int_\tau^t dt' h^-(t') \right\} \tag{18}$$

$$g^-(t) = \int d\vec{r}\, \varphi_f^{-*}(\vec{r}) \exp\left( -i\vec{A}^-(t).\vec{r} \right) \varphi_i(\vec{r}) \tag{19}$$

With the expression (7) of the the external field $\vec{F}(t)$, the functions $h^-(t)$ and $f^-(t)$ may be calculated analytically. If the form of $\vec{F}(t)$ is more complicated, it is not difficult to perform accurate numerical time integrations. One may also get an analytical expression for $g^-(t)$ using a standard procedure [10]. According to (10), one has :

$$\frac{\partial}{\partial t} g^-(t) = i \int d\vec{r}\, \varphi_f^{-*}(\vec{r}) \exp\left( -i\vec{A}^-(t).\vec{r} \right) \vec{r}.\vec{F}(t)\, \varphi_i(\vec{r}) \tag{20}$$

Thus, $T_{fi}^{CV2-}$ may be written as :

$$T_{fi}^{CV2-} = -\int_0^\tau dt\, f^-(t) \frac{\partial}{\partial t} g^-(t) \qquad (21)$$

Integrating by parts and bearing in mind that $\vec{A}^-(\tau) = \vec{0}$ , one obtains :

$$T_{fi}^{CV2-} = f^-(0)\, g^-(0) - \int_0^\tau dt\;\; h^-(t)\; f^-(t)\, g^-(t) \qquad (22)$$

It is worth noting that the first term of the right -hand side in (22) is nothing but $T_{fi}^{CV1-}$ (prior version of C V1) multiplied by the phase factor $f^-(0)$ [4,6]. For a genuine laser pulse this term is zero, since one has also $\vec{A}^-(0) = \vec{0}$ (no direct electric field). Therefore, a simple numerical time integration over the pulse length is ne cessary to know $T_{fi}^{CV2-}$. Then, the angular distribution of ejected electrons is given by :

$$\frac{\partial^2 P_{fi}^{CV2-}}{\partial E_K\, \partial \Omega_k} = k \left| T_{fi}^{CV2-} \right|^2 \qquad (23)$$

where $E_k$ and $\Omega_k$ are the energy and the direction corresponding to the impulse $\vec{k}$ of an ejected electron. Integrating over $\Omega_k$ gives the energy distribution $\frac{\partial P_{fi}^{CV2-}}{\partial E_K}$ and a further integration over $E_k$ gives the total probability $P_{fi}^{CV2-}$ to ion ize an atom with one pulse. *A priori*, one expects good predictions from the present treatment as long as $P_{fi}^{CV2-} << 1$. The aim of the following section is to determine the upper acceptable value of $P_{fi}^{CV2-}$.

## III. Application of CV2$^-$ to ionization of hydrogen atoms

To determine under which conditions CV2$^-$ applies, we address here the ionization of hydrogen atoms in their ground state by high -frequency laser pulses. As already mentioned, the study is first carried out with a photo n energy $\omega = 0.855$ (15$^{th}$ harmonics in ref. [1] ) in order to connect it to a more realistic case. Then, we investigate $P_{fi}^{CV2-}$ as a function of $\omega$ with laser parameters that ensure $P_{fi}^{CV2-} << 1$ whatever $\omega$.

### A. Influence of the laser intensity

Let us first set two laser pulse parameters : $\omega = 0.855$ and $\tau = 100$ . Thus $\tau$ ( 2.4 $fs$ ) is comparable to the duration of a *single* high-harmonic pulse [12], and it permits to get well separated ATI peaks. Further, a single photon absorption is enough to ionize $H(1s)$. Electron energy spectra as predicted by CV2$^-$ and TDSE are reported on Fig. 1 for increasing laser intensities. For laser field amplitudes up to $F_0 = 0.16$ $\left(I \approx 9\times10^{14} W.cm^{-2}\right)$, there is an excellent agreement between the two spectra. The shape of peaks are very well reproduced by CV2$^-$. The difference between CV2$^-$ and TDSE backgrounds that shows up in between two consecutive peaks at the highest ejection energies for $F_0 \geq 0.04$, may be due, not to a shortcoming of the CV2$^-$ method itself, but to rounding-off numbers in the numerical time integration. For the background, the contribution of a given half-cycle is both, very small compared to the value at a peak and very close to the opposite of the contribution of the subsequent half-cycle. Thus, the final value seems mainly connected to the order of magnitude of the last significant digit.

Now significant differences show up for $F_0 = 0.32$, i.e., $I \approx 3.6\times10^{15} W.cm^{-2}$ (Fig.1f). All ATI peaks predicted by CV2$^-$ are both, higher than TDSE ones, and all shifted towards the ionization threshold by the same value. In principle, due to energy conservation, the ATI peaks appear at energies verifying :

$$E_n = \varepsilon_i + n\omega - U_p \tag{24}$$

where $n$ is the number of absorbed photons and where $U_p$ is a ponderomotive energy. In perturbation conditions, one has $U_p = \dfrac{F_0^2}{4\omega^2}$, $F_0^2$ being the peak intensity. However, in the very case where $F_0 = 0.32$ , the total ionization probability given by TDSE is $P_{fi}^{TDSE} = 0.72$, which indicates that ionisation saturates. It means that the atom looses the opportunity of experiencing the maximum intensity since the electron has gone before. Rather, ionization occured at some effective intensity during the pulse raise when $U_p$ was smaller. The effect is clearly reproduced in TDSE calculations. At variance, CV2$^-$ does not account for saturation, as it is indicated by its unrealistic prediction of the total ionization probability, i.e., $P_{fi}^{CV2-} = 1.44$. In fact, CV2$^-$ predicts that $\dfrac{\partial P_{fi}}{\partial E_K}$ is maximum when the electric field amplitude is maximum, i.e., when $F(t) = F_0$. In addition, CV2$^-$ overestimates $\frac{\partial P_{fi}}{\partial E_K}$ in the vicinity of $F_0$, for $F_0 = 0.32$, the closer to $F_0$, the higher the overestimation. Therefore, the maxi mum of a ATI peak given by CV2$^-$ shows up when the ponderomotive energy reaches its highest value, that is $U_p \approx 3.5\times10^{-2}$ for $F_0 = 0.32$. Since TDSE never overestimates $\frac{\partial P_{fi}}{\partial E_K}$, one expects the shift to amount to this maximum value of the ponderomotive energy. It is what is observed on Fig.1f. In fact, a similar (but very tiny) shift may be detected on Fig.1e. If saturation conditions still prevailed, it would be $9\times10^{-3}$ according to formula (24). Since it is not the case, the shift is even smaller, thus being almost not visible.

On Fig.2, total ionization probabilities $P_{fi}^{CV2-}$ and $P_{fi}^{TDSE}$ are plotted as functions of the laser intensity $I = F_0^{\,2}$. The two curves stay close to each other till $P_{fi}^{CV2-}$ approaches 1. They may be distinguished as soon as $I \geq 10^{-3}$, but it is worth noting that, even for values of $P_{fi}^{CV2-}$ greater than 10%, CV2$^-$ and TDSE energy distributions are still in good agreement. For instance, when $I = 2.6\times10^{-2}$ $(F_0 = 0.16)$, one has $P_{fi}^{CV2-} \approx 0.33$ while $P_{fi}^{TDSE} \approx 0.28$, but the CV2$^-$ energy distribution still agrees with TDSE. Positions and heights of the first five peaks are pretty much the same. Therefore, CV2$^-$ works in a domain that stretches slightly beyond the perturbation regime.

### B. Influence of the pulse duration

In Fig.3, we set $F_0 = 0.05$ so that one can approach the limits of the perturbation regime by increasing $\tau$. Calculations are performed for $\omega = 0.855$ and for $\tau$ ranging from 20 to 500. The situation looks very similar to the above section III.A. A good agreement is found everywhere for ATI peaks except for the last two peaks when $\tau = 500$. Again, due to the loss of significant digits in between peaks, CV2$^-$ predictions for the background are all the worse that $\tau$ is large. However, even for $\tau = 500$, the total ionization probabilities given by the two approaches are the same ($P_{fi}^{CV2-} = P_{fi}^{TDSE} = 0.155$). Keeping in mind that high-harmonic pulses last 200 a.u. at most, it is a good indication that CV2$^-$ is a valuable tool to address realistic cases.

## C. Influence of the photon energy

So far, we have examined electron energy spectra that are obtained with a photon energy greater than the ionization threshold. Let us now set the laser parameters $F_0 = 0.01$ and $\tau = 100$ to ensure that the total probability stays well below 1 whatever $\omega$. Therefore, the study is always made in the perturbation regime. In Fig.4, we compare $P_{fi}^{CV2-}$ to $P_{fi}^{TDSE}$ while $\omega$ is increased from 0 to 1. Both predictions cannot be distinguished above $\omega = 0.5$, i.e., with a photon energy greater than the ionization potential. A reasonable agreement is still found for $0.42 \le \omega \le 0.5$ due to the wide broadening of the pulse frequency for $\tau = 100$. The two predictions disagree by an order of magnitude between 0 and 0.42. A rough qualitative explanation of this behaviour may be given as follows : $L^-(\vec{r},t)$ contains the displacement factor $\exp\left\{ -i\vec{k}\cdot\int_\tau^t dt'\vec{A}^-(t') \right\}$, thus leadind to a factor $\exp\left\{ i\vec{k}\cdot\left[ \vec{r} - \int_\tau^t dt'\vec{A}^-(t')\right] \right\}$ in $\chi_f^-(\vec{r},t)$. The second term in the square-bracket is nothing but the classical displacement of a free electron under the influence of the laser electric field during the pulse. In the case $\omega \ge 0.5$, the ejected electron is "free to move immediately" in the continuum after the absorption of any number of photons. However, when $\omega < 0.5$ two or more photons are necessary to achieve ionization. Thus, although the electron is not "immediatly in the continuum", CV2$^-$ anticipates its displacement that is equivalent to anticipating ionization. As a result, ionization is overestimated. An other cause of disagreement comes from the fact that CV2$^-$ does not account for intermediate Rydberg resonances when $\omega \le 0.5$.

Therefore, the present form of CV2$^-$ provides reliable data only when the photon energy is greater than the ionization potential. It is worth noting that similar calculations have been performed using the post form of the transition amplitude in which the Coulomb-Volkov state $\chi_i^+(\vec{r},t)$ connected to the initial states replaces $\Psi_i^+(\vec{r},t)$ [4,6]. Since the electron displacement is missing in this approach, called CV2$^+$, predictions are much worse than CV2$^-$ ones [11].

## D. Application in the collisional regime

In all cases that have been examined in previous paragraphs, the electric field of the laser performs about 3 oscillations or more, thus placing the transition in the "photonic regime". Although there is nothing indicating that CV2$^-$ could fail in the "collisional regime" (where the electromagnetic field performs less than 2 oscillations), it is worth checking that good predictions can still be obtained. To do that, we kept the first term in the r.h.s. of Eq. (22) since one has generally $\vec{A}^{-}(0) \neq \vec{0}$ in this case. On Fig. 5, electron spectra are reported for $F_0 = 0.01$ and $\omega = 0.855$. The pulse duration is increased from $\tau = 5$ (collisional regime with less than one oscillation) to $\tau = 100$ (photonic regime with almost 14 oscillations). Predictions for $\tau = 20$ are also displayed because the situation corresponds to the parameters of Paul *et al.*'s experiment [1]. As expected, $CV2^-$ agrees well with TDSE in all cases.

## IV. Conclusions and perspectives

Atom ionization by short VUV laser pulses may be described accurately by the first order of a perturbation approach based on Coulomb-Volkov states. This approach, that we called $CV2^-$, consists in replacing, in the prior form of the transition amplitude, the total exact wavefunction connected to the final continuum state by the corresponding Coulomb-Volkov wavefunction. On the classic example of atomic hydrogen targets, we have shown that very good predictions of ATI spectra are obtained when both, the photon energy is greater than or equal to the ionization potential of the target, and perturbation conditions prevail. In the present case, our study shows that the total ionization probability should not exceed, let us say 20% to 30% to get accurate electron spectra. Therefore, $CV2^-$ looks very promising to study the ionization of atoms or molecules by the high-harmonics laser pulses which are now generated.

The present study is the first extended test of $CV2^-$ with subfemtosecond extreme ultraviolet laser pulses. Some defects showed up in the calculation of the spectrum background. Hence, we plan to improve the time integration procedure in order to extend the domain where the method $CV2^-$ applies.

**Figure captions :**

Figure 1 : ionization of $H(1s)$ : electron distribution (density of probability per energy range) as a function of the energy of the ejected electron for a photon energy $\omega = 0.855$, a pulse length $\tau = 100$ and various laser field amplitudes $F_0$. (a) $F_0 = 0.01$ ; (b) $F_0 = 0.02$ ; (c) $F_0 = 0.04$ ; (d) $F_0 = 0.08$ ; (e) $F_0 = 0.16$ ; (f) $F_0 = 0.32$. All quantities are given in atomic units. Dotted line : $CV2^-$ ; full line : TDSE.

Figure 2 : total ionization probability of $H(1s)$ as a function of the laser intensity $I = F_0^2$ for a photon energy $\omega = 0.855$ and a pulse length $\tau = 100$. All quantities are given in atomic units. Dotted line : $CV2^-$ ; full line : TDSE.

Figure 3 : ionization of $H(1s)$ : electron distribution as a function of the energy of the ejected electron for a photon energy $\omega = 0.855$, a laser field amplitude $F_0 = 0.05$ and various pulse lengths. (a) $\tau = 20$ ; (b) $\tau = 50$ ; (c) $\tau = 100$ ; (d) $\tau = 200$ ; (e) $\tau = 350$ ; (f) $\tau = 500$. All quantities are given in atomic units. Dotted line : $CV2^-$ ; full line : TDSE.

Figure 4 : total ionization probability of $H(1s)$ as a function of the photon energy for a laser field amplitude $F_0 = 0.01$ and a pulse length $\tau = 100$. All quantities are given in atomic units. Dotted line : $CV2^-$ ; full line : TDSE.

Figure 5 : ionization of $H(1s)$ : electron distribution as a function of the energy of the ejected electron for a photon energy $\omega = 0.855$, a laser field amplitude $F_0 = 0.01$ and three pulse lengths $\tau$. (a) $\tau = 5$ (collisional regime); (b) $\tau = 20$ (lower limit of the photonic regime); (c) $\tau = 100$ (photonic regime). All quantities are given in atomi c units. Dotted line : $CV2^-$ ; full line : TDSE.

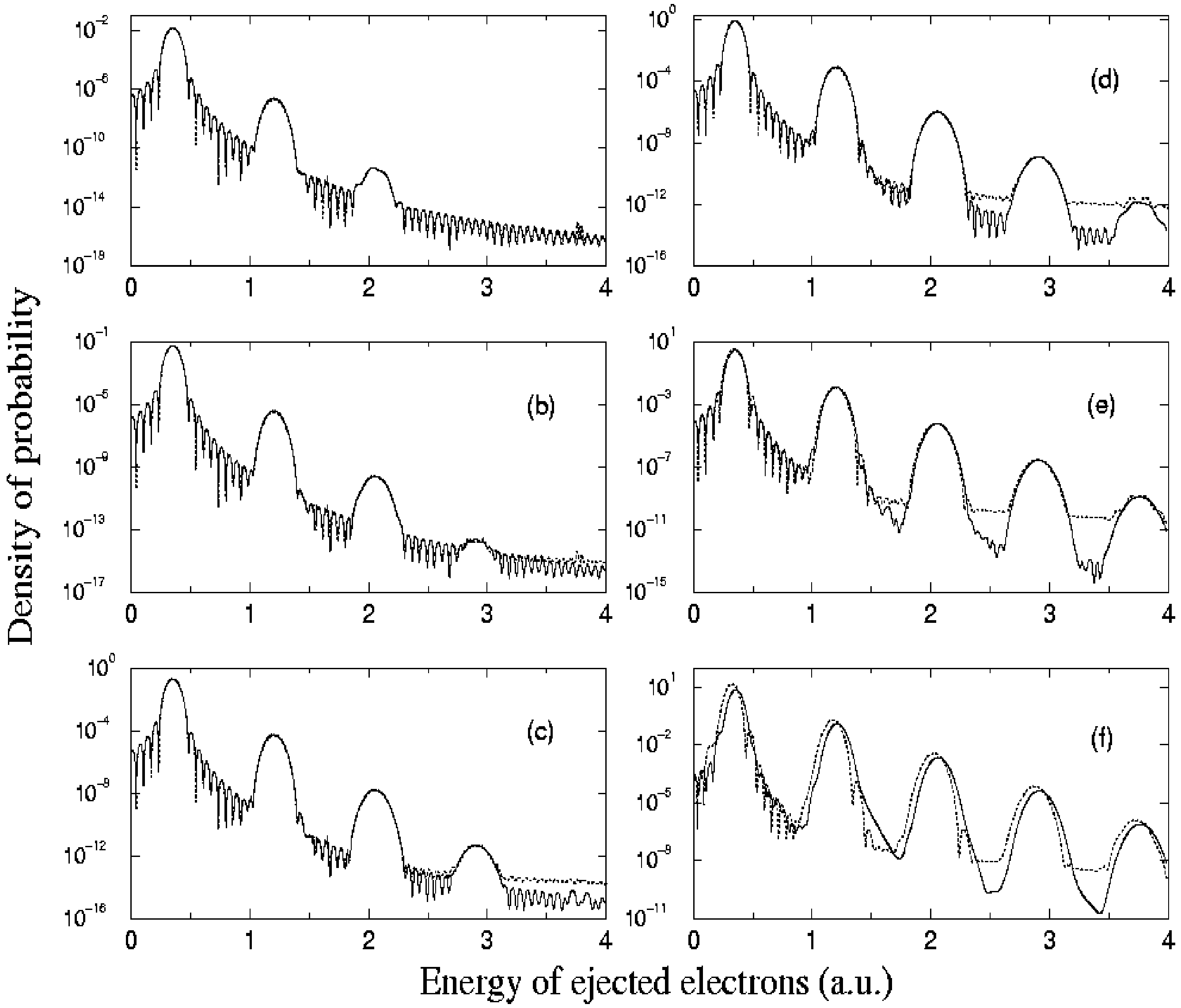

**Figure 1**

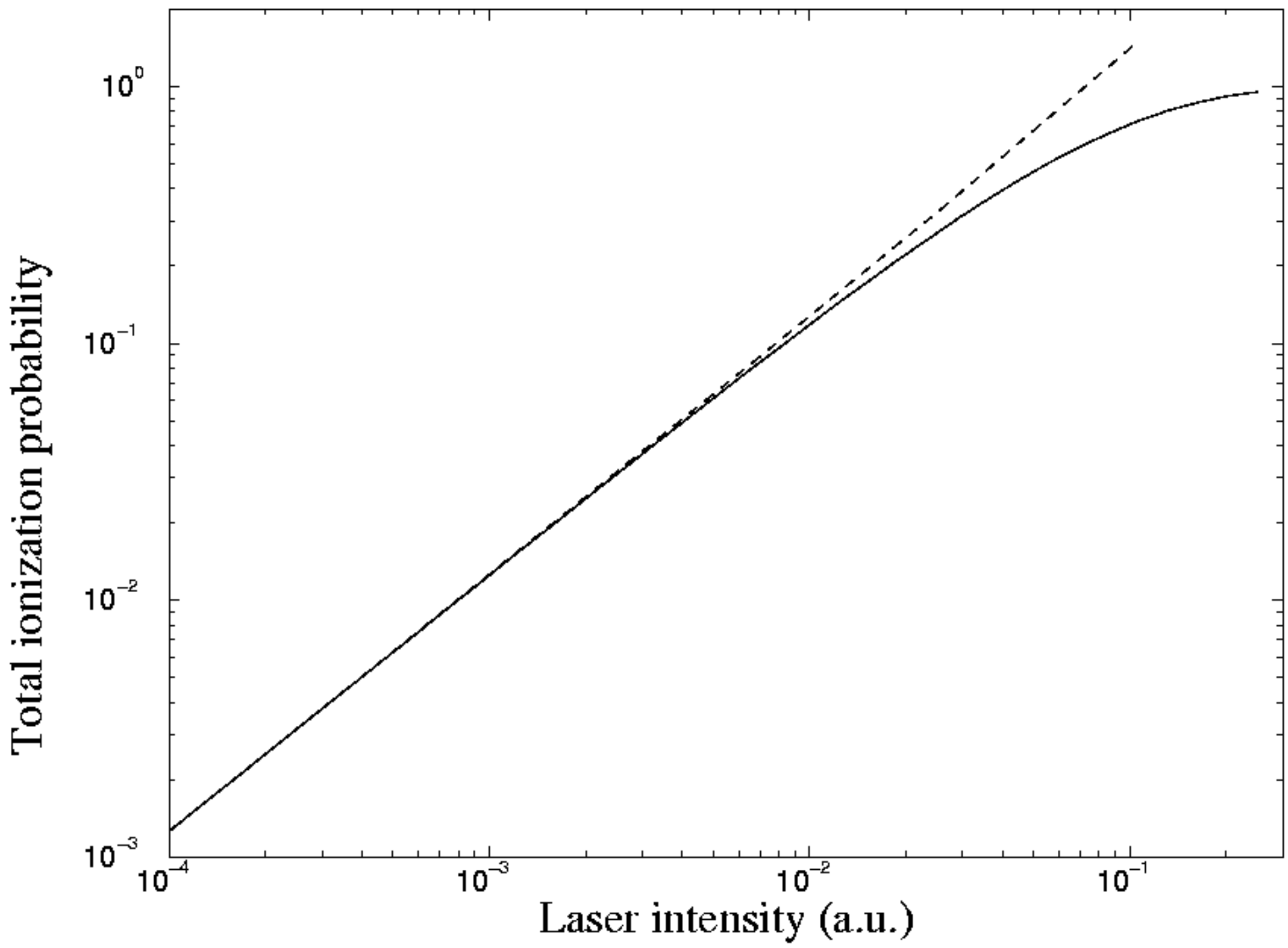

**Figure 2**

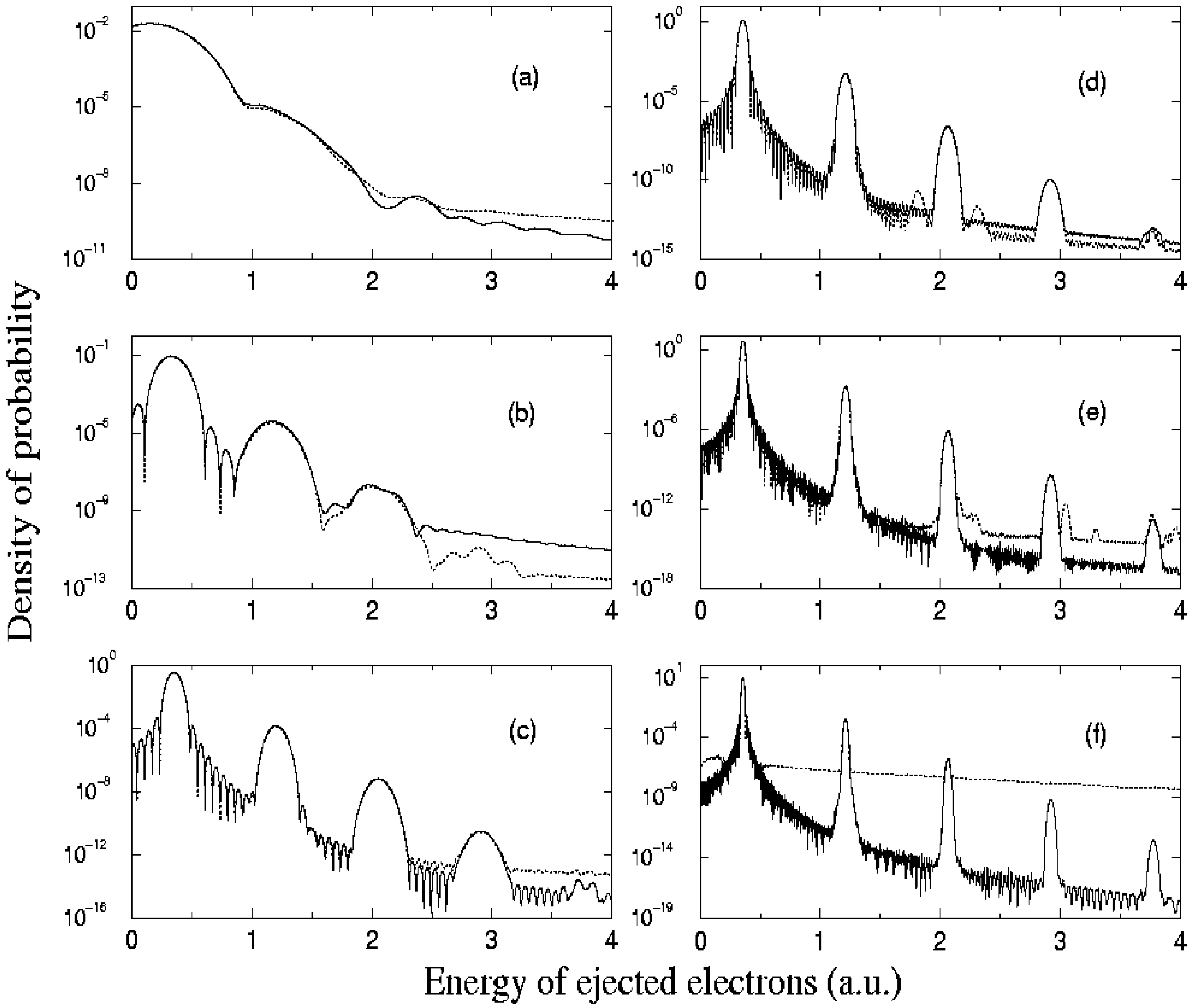

**Figure 3**

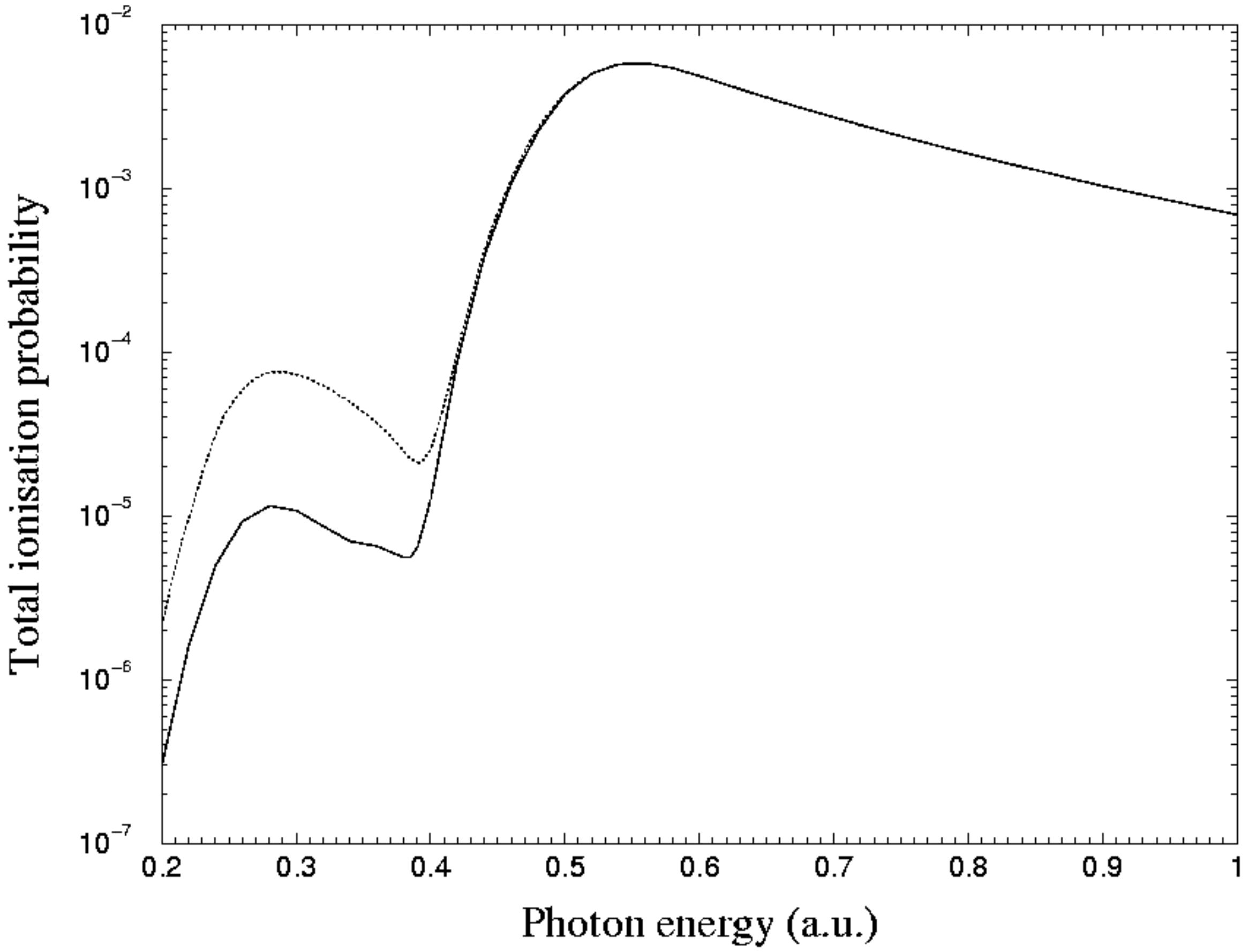

**Figure 4**

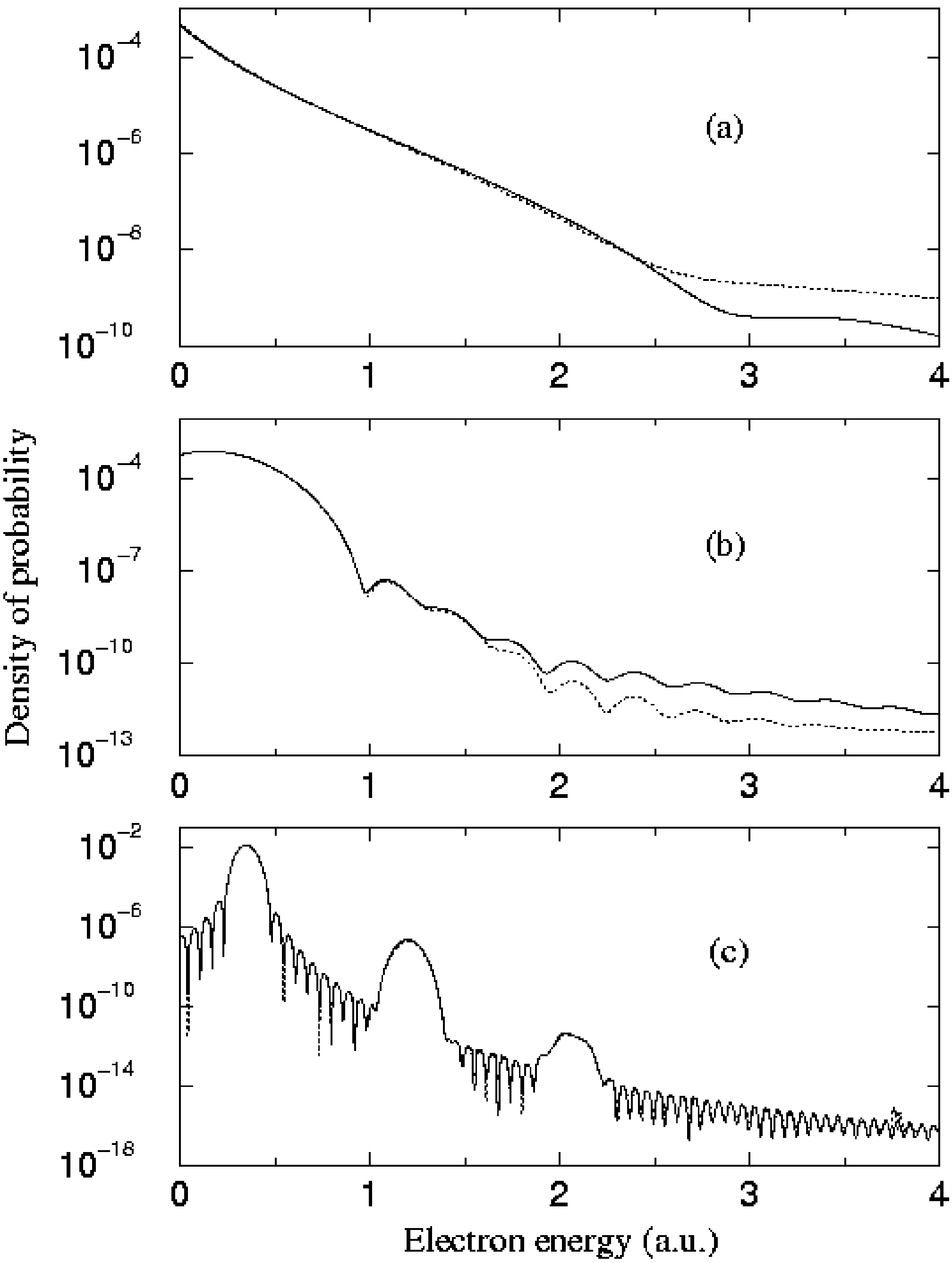

**Figure 5**